\documentclass[useAMS]{mn2e}
\usepackage{aas_symbols}
\usepackage{times}
\usepackage{rotating}
\usepackage{multirow}
\usepackage{cancel}
\long\def\symbolfootnote[#1]#2{\begingroup%
\def\thefootnote{\fnsymbol{footnote}}\footnote[#1]{#2}\endgroup}
\def\G{\Gamma}
\def\g{\gamma}
\def\b{\beta}
\def\al{\alpha}
\begin{document}

\title[Prompt Optical and Site of GRBs]{Prompt optical emission and synchrotron self-absorption constraints on emission site of GRBs}

\author[Shen \& Zhang]{Rong-Feng Shen$^{1}$\thanks{E-mail: rfshen@astro.as.utexas.edu (R-FS); zhang@physics.unlv.edu (BZ)} and Bing Zhang$^{2}$\footnotemark[1]\\
$^{1}$Department of Astronomy, University of Texas at Austin, Austin, TX 78712, USA\\
$^{2}$Department of Physics and Astronomy, University of Nevada at Las Vegas, Las Vegas, NV 89154, USA}

\date{Accepted 2009 June 06. Received 2009 June 06; in original form 2008 July 23}

\pagerange{\pageref{000}--\pageref{000}} \pubyear{2008}

\maketitle

%%%%%%%%%%%%%%%%%%%%%%%%%%%%%%%%%%%%%%%%%%%%%%%%%%%%%%%%%%%%%%%%%%%%%%%%%%%%%%%%%%%%
\begin{abstract}
We constrain the distance of the Gamma-Ray Burst (GRB) prompt emission site from the explosion centre, $R$, by determining the location of the electron's self absorption frequency in the GRB prompt optical-to-X/$\g$-ray spectral energy distribution, assuming that the optical and the $\g$-ray emissions are among the same synchrotron radiation continuum of a group of hot electrons. All possible spectral regimes are considered in our analysis. The method has only two assumed parameters, namely, the bulk Lorentz factor of the emitting source $\G$, and the magnetic field strength $B$ in the emission region (with a weak dependence). We identify a small sample of 4 bursts that satisfy the following three criteria:
(1) they all have simultaneous optical and $\g$-ray detections in multiple observational time intervals; (2) they all show temporal correlations between the optical and $\g$-ray light curves; and (3) the optical emission is consistent with belonging to the same spectral component as the $\g$-ray emission. For all the time intervals of these 4 bursts, it is inferred that $R \geq 10^{14} \,\, (\G/300)^{3/4} (B/10^5 \, {\rm G})^{1/4}$ cm.
For a small fraction of the sample, the constraint can be pinned down to $R \approx 10^{14} - 10^{15}$ cm for $\G \sim 300$. For a second sample of bursts with prompt optical non-detections, only upper limits on $R$ can be obtained. We find no inconsistency between the $R$-constraints for this non-detection sample and those for the detection sample.
\end{abstract}


%%%%%%%%%%%%%%%%%%%%%%%%%%%%%%%%%%%%%%%%%%%%%%%%%%%%%%%%%%%%%%%%%%%%%%%%%%%%%%%%%%%%%

%%%%%%%%%%%%%%%%%%%%%%%%%%%%%%%%%%%%%%%%%%%%%%%%%%%%%%%%%%%%%%%%%%%%%%%%%%%%%%%%%
\section{Introduction}

Although the Gamma-Ray Burst (GRB) was discovered about 50 years ago first through its prompt $\g$-ray emission, large uncertainties still remain in understanding the prompt emission site, namely, the distance of the emission site from the explosion centre $R$, with controversial evidence. There are three possible sites discussed in the literature. One is the standard internal-shock site which depends on the fluctuation time scale $\delta t$ seen in GRB light curves (e.g., Rees \& M\'{e}sz\'{a}ros 1994, see Piran 2005, M\'esz\'aros 2006 for reviews). It can have a large range of $R \sim \G^2 c \delta t \sim 10^{13}-10^{15}$ cm because $\delta t$ and $\G$ vary largely from burst to burst. The second is the photospheric radius at $10^{11}-10^{12}$ cm at which the prompt emission arises as a combination of the photosphere thermal emission and a Comptonized component above it, the latter being induced by some energy dissipation process below and above the photosphere (e.g. Rees \& M\'{e}sz\'{a}ros 2005; Ryde et al. 2006; Thompson et al. 2007). The third one is a large radius ($> 10^{14}$ cm) as is supported by the Swift XRT data (Lazzati \& Begelman 2005; Lyutikov 2006; Kumar et al. 2007) and Fermi data of GRB 080916C (Abdo et al. 2009; Zhang \& Pe'er 2009), possibly due to magnetic dissipation (e.g., Lyutikov \& Blandford 2003).

The rapidly responding ability of a few GRB-dedicated ground or space based optical telescopes, e.g., ROTSE (Akerlof et al. 2003), RAPTOR (Vestrand et al. 2002), TORTORA (Racusin et al. 2008) and the UVOT (Roming et al. 2005) on aboard the Swift satellite, has enabled the time-resolved detection of bright prompt optical emission before the $\g$-rays die off, for about a dozen of GRBs. Five of these GRBs, i.e., 041219A (Vestrand et al. 2005), 050820A (Vestrand et al. 2006), 051111 (Yost et al. 2007a), 061121 (Page et al. 2007) and 080319B (Racusin et al. 2008), show a temporal correlation between the strongly variable optical flux and the $\g$-ray pulses, which suggests that the optical emission most likely shares the same dynamical process that is responsible for the highly variable $\g$-ray emission. While the other four bursts have optical flux densities below or marginally consistent with the extrapolations from the low-energy power law of the $\g$-ray spectra, the optical flux density in GRB 080319B exceeds the $\g$-ray extrapolation by 4 orders of magnitude (Racusin et al. 2008; Kumar \& Panaitescu 2008), suggesting that for this burst alone the optical emission has a spectral origin different from that of the $\g$-rays.

In this paper, for the four GRBs - 041219A, 050820A, 051111 and 061121 - we assume that the prompt optical and the $\g$-ray emissions are components belonging to the same synchrotron radiation continuum of a group of hot electrons. Based on this assumption, the self-absorption frequency of the synchrotron electrons, $\nu_a$, which causes a break in the long-wavelength part of the continuum, can be determined or constrained by studying the optical-to-$\g$-ray spectral energy distribution (SED)\footnote{The significance of self-absorption frequency has been highlighted by Doi, Takami \& Yamazaki (2007) who used the varying location of $\nu_a$ to interpret the diversity in the prompt optical / $\g$-ray temporal correspondence.}. Since $\nu_a$ is dependent on the properties of the prompt emission source, such as the distance of the emission site from the explosion centre $R$, the bulk Lorentz factor (LF) $\G$ and the magnetic field $B$ of the source, from $\nu_a$ we can determine or make constraints on $R$ for these bursts, using information on $\G$ and $B$ obtained in other ways. This is the main goal of this paper. Since the prompt optical and $\g$-ray components in GRB 080319B are most likely of different spectral origins because of its peculiar SED shape, our approach is not applicable to this burst.

On the other hand, for some other long GRBs the rapid response of the dedicated ROTSE telescope has returned only upper limits of the optical flux density during the prompt phase (Yost et al. 2007b). Another goal of this paper is to get constraints on $R$ for these optically ``dark'' bursts and to study whether the prompt optical non-detection is caused by a heavier self-absorption due to a closer emission site to the explosion centre.

In this paper, we first derive analytically $\nu_a$ in terms of $R$, $\G$, $B$ and the emission properties in Sec. 2. The arguments that support our assumption of one synchrotron continuum component for both optical and $\g$/X-ray are given in Sec. 3.  We derive in Sec. 4 the constraints on $R$ through $\nu_a$ explicitly, by determining the location of $\nu_a$ in the optical-to-$\g$-ray SED and considering all possible spectral regimes. We apply this method to a prompt optical detection GRB sample and a prompt optical non-detection sample which are described in Sec. 5. The results are presented in Sec. 6. Finally the conclusion and discussions are given in Sec. 7.

%%%%%%%%%%%%%%%%%%%%%%%%%%%%%%%%%%%%%%%%%%%%%%%%%%%%%%%%%%%%%%%%
\section{Determining the self-absorption frequency}

The GRB high energy emission spectrum is characterized by a smoothly joint broken-power-law form (Band et al. 1993). Thus the relativistic electrons responsible for the GRB prompt emission due to either synchrotron or synchrotron self-inverse Compton (SSC) radiation are in a piece-wise two-power-law energy distribution:

\begin{equation}
N(\g) \propto \left\{ \begin{array}{ll}
\g^{-p_1}, & {\rm if}\,\, \g_m < \g < \g_p \,,\\
\g^{-p_2}, & {\rm if}\,\, \g > \g_p \,,\\
\end{array} \right.
\end{equation}
where $N(\g)$ is such defined that $N(\g)d\g$ is the number density of
electrons with energy in the interval of $\g$ to ($\g+d\g$), and $\g_m$ is
the minimum energy of these relativistic electrons (for convenience we omit the factor $m_e c^2$ in the electron energy $\g m_e c^2$ throughout the text when electron energy is mentioned).

Note that although this distribution set-up is phenomenologically based on the two-power-law shape of the high energy radiation spectrum observed in GRBs, it has specific physical meanings. Within
the shock acceleration scenario, newly accelerated electrons with a minimum energy $\g_i$ and a power-law energy distribution are continuously injected. These electrons lose energy through radiative cooling, and the instantaneous electron spectrum steepens above a critical energy $\g_c$.
All the electrons with energy larger than $\g_c$ radiate away their energy within a time shorter than the dynamical time. When $\g_i < \g_c$, our notation corresponds to $\g_m= \g_i$ and $\g_p=\g_c$. When $\g_c < \g_i$, the cooling causes a flatter power law between $\g_c$ and $\g_i$, even though the newly accelerated electrons are injected in the energy range above $\g_i$. For this case one has $\g_m= \g_c$ and $\g_p= \g_i$. In summary within the shock acceleration scenario one has $\g_m=\min(\g_i, \g_c)$ and $\g_p=\max(\g_i, \g_c)$. More generally, one can also have a scenario that invokes continuous heating and cooling of electrons (e.g. that envisaged in the reconnection models), and $\g_p$ then reflects the intrinsic break in the steady state electron spectrum. In any case, our treatment is generic, which does not depend on the concrete particle acceleration mechanism and the origin of $\g_p$.

The broken power-law electron energy spectrum naturally gives rise to a piece-wise power law photon spectrum as commonly observed:
\begin{equation}
f_{\nu} \propto \left\{ \begin{array}{ll}
\nu^{\b_1}, & {\rm if}\,\, \nu_m < \nu < \nu_p \,,\\
\nu^{\b_2}, & {\rm if}\,\, \nu > \nu_p \,,\\
\end{array} \right.
\end{equation}
where $f_{\nu}$ is the observed flux density (in units of mJy), $\nu_m$ and $\nu_p$ are the observed characteristic emission frequencies of the electrons with energy $\g_m$ and $\g_p$, respectively, and $\nu_p$ is usually the peak frequency of the $\nu f_{\nu}$ spectrum. The low-energy power law $\nu^{\b_1}$ does not extend to low frequency indefinitely. Without synchrotron self absorption, the spectral index below $\nu_m$ changes to 1/3, regardless of whether $\g_i < \g_c$ or $\g_c < \g_i$. Below a certain frequency $\nu_a \ll \nu_p$, the synchrotron self absorption starts to play a significant role -- at frequencies lower than $\nu_a$ the emitted photons are thermalized with the electrons. The self-absorption frequency $\nu_a$ is such defined that at this frequency the self-absorption optical depth $\tau_{sa}(\nu_a)= 1$.

Let us determine $\nu_a$ for an emitting GRB ejecta moving with a Lorentz factor (LF) $\G$ at a distance $R$ from the center of the explosion. In the ejecta comoving frame (hereafter the quantities measured in this frame are marked with the prime sign), $\nu_a'$ can be determined by equating the un-absorbed source surface flux density, $F_{\nu_a'}'$, at $\nu_a'$ to a blackbody surface flux density with temperature $T'$ in the Rayleigh-Jeans regime (see Appendix for the derivation; also see Sari \& Piran 1999, Li \& Song 2004, McMahon, Kumar \& Piran 2006):
\begin{displaymath}
2 k T' \frac{\nu_a'^2}{c^2} = C(\b_1) F_{\nu_a'}' \\
\end{displaymath}
\begin{equation}
=\left\{ \begin{array}{ll}
C_1(\b_1) F_{\nu_p'}' \left(\frac{\nu_m'}{\nu_p'}\right)^{\b_1} \left(\frac{\nu_a'}{\nu_m'}\right)^{\frac{1}{3}}, & {\rm for\,\,\, } \nu_a' < \nu_m',\\
C_2(\b_1) F_{\nu_p'}' \left(\frac{\nu_a'}{\nu_p'}\right)^{\b_1}, & {\rm for}\,\,\, \nu_m' < \nu_a' < \nu_p', \\
\end{array} \right.
\end{equation}
where $T'= \max(\g_a, \g_m) m_e c^2/k$, $\g_a$ is the energy of electron whose characteristic emission frequency is $\nu_a'$, and $k$ is the Boltzmann constant. The numerical factors $C_1$ and $C_2$ are functions of $\b_1$ only whose values range from 1.2 to 4.5 and from 1.2 to 7.0, respectively, for the range of observed $\b_1$ values, but they have been neglected in previous works while we include them here.

Transforming the frequency to that measured in the observer's frame gives $\nu_a= \nu_a' \G/(1+z)$, where $z$ is the redshift of the GRB host galaxy. Measuring in the host comoving frame, the source has an isotropic luminosity of $4\pi R^2 \G^2 F_{\nu_p'}'\nu_p'$. This luminosity is also given by $4\pi D_L^2 f_{\nu_p}\nu_p$, where $D_L$ is the luminosity distance of the GRB and $f_{\nu_p}$ is the observed peak flux density. Thus we have $F_{\nu_p'}'= f_{\nu_p}(D_L/R)^2/[\G(1+z)]$. After applying these relations, Eq. (3) becomes
\begin{equation}
\left\{ \begin{array}{ll}
\frac{C_1}{2} f_{\nu_p} \left(\frac{\nu_m}{\nu_p}\right)^{\b_1} \left(\frac{\nu_a}{\nu_m}\right)^{\frac{1}{3}} & =
m_e \g_m \nu_a^2 \left(\frac{R}{D_L}\right)^2 \frac{(1+z)^3}{\G},\\
& {\rm \quad \ for\ } \nu_a < \nu_m; \\
\frac{C_2}{2} f_{\nu_p} \left(\frac{\nu_a}{\nu_p}\right)^{\b_1} & = m_e \g_a \nu_a^2 \left(\frac{R}{D_L}\right)^2 \frac{(1+z)^3}{\G},\\
& {\rm \quad \ for\ } \nu_m < \nu_a. \\
\end{array} \right.
\end{equation}

After substituting the electron's energy $\g$ using the following relation between $\g$ and
the photon frequency $\nu$,
\begin{equation}
\g = \left\{ \begin{array}{ll}
\left(\frac{2\pi m_e c}{eB}\right)^{\frac{1}{2}} \left(\frac{1+z}{\G}\right)^{\frac{1}{2}} \nu^{\frac{1}{2}},
& {\rm \  for \ synchrotron},\\
\left(\frac{2\pi m_e c}{eB}\right)^{\frac{1}{4}} \left(\frac{1+z}{\G}\right)^{\frac{1}{4}} \nu^{\frac{1}{4}},
& {\rm \  for \ SSC},\\
\end{array} \right.
\end{equation}
the self-absorption frequency is calculated as: for synchrotron
\begin{displaymath} \nu_a = \end{displaymath}
\begin{equation}
\left\{ \begin{array}{l}
\left(\frac{C_1}{2}\right)^{\frac{3}{5}}\times10^{14.6-\frac{6}{5}\b_1} f_{\nu_p}^{\frac{3}{5}} \nu_{p,19}^{-\frac{3}{5}\b_1} \nu_{m,17}^{\frac{3}{5}\b_1-\frac{1}{2}}\\
\hspace{1.5cm} \times \bigl(\frac{D_{L,28}}{1+z}\bigr)^{\frac{6}{5}} \bigl(\frac{\G_{300}}{1+z} \bigr)^{\frac{9}{10}}
B_5^{\frac{3}{10}} R_{14}^{-\frac{6}{5}} {\rm \,\,Hz}\\
\hspace{4.5cm} {\rm if}\,\, \nu_a < \nu_m;\\
\left(\frac{C_2}{2}\right)^{\frac{1}{2.5-\b_1}}\times10^{\frac{38.5-19\b_1}{2.5-\b_1}}
f_{\nu_p}^{\frac{1}{2.5-\b_1}} \nu_{p,19}^{\frac{-\b_1}{2.5-\b_1}}\\
\hspace{0.5cm} \times \bigl(\frac{D_{L,28}}{1+z}\bigr)^{\frac{2}{2.5-\b_1}} \bigl( \frac{\G_{300}}{1+z} \bigr)^{\frac{1.5}{2.5-\b_1}} B_5^{\frac{1}{5-2\b_1}}
R_{14}^{\frac{-2}{2.5-\b_1}} {\rm \,\,Hz}\\
\hspace{4.5cm} {\rm if}\,\, \nu_m < \nu_a < \nu_p\\
\end{array} \right.
\end{equation}
and, for SSC
\begin{displaymath} \nu_a = \end{displaymath}
\begin{equation}
\left\{ \begin{array}{l}
\left(\frac{C_1}{2}\right)^{\frac{3}{5}}\times10^{15-\frac{6}{5}\b_1} f_{\nu_p}^{\frac{3}{5}} \nu_{p,19}^{-\frac{3}{5}\b_1}
\nu_{m,17}^{\frac{3}{5}\b_1-\frac{7}{20}}\\
\hspace{1.5cm} \times \bigl(\frac{D_{L,28}}{1+z}\bigr)^{\frac{6}{5}} \bigl( \frac{\G_{300}}{1+z} \bigr)^{\frac{3}{4}} B_5^{\frac{3}{20}} R_{14}^{-\frac{6}{5}} {\,\,\rm Hz} \\
\hspace{4.5cm} {\rm if}\,\, \nu_a < \nu_m; \\
\left(\frac{C_2}{2}\right)^{\frac{1}{2.25-\b_1}}\times10^{\frac{35-19\b_1}{2.25-\b_1}}
f_{\nu_p}^{\frac{1}{2.25-\b_1}} \nu_{p,19}^{\frac{-\b_1}{2.25-\b_1}} \\
\hspace{0.3cm} \times \bigl(\frac{D_{L,28}}{1+z}\bigr)^{\frac{2}{2.25-\b_1}} \bigl(\frac{\G_{300}}{1+z} \bigr)^{\frac{1.25}{2.25-\b_1}} B_5^{\frac{1}{9-4\b_1}} R_{14}^{\frac{-2}{2.25-\b_1}} {\,\,\rm Hz}\\
\hspace{4.5cm} {\rm if}\,\, \nu_m < \nu_a < \nu_p \,.\\
\end{array} \right.
\end{equation}
In the results above, the flux density, e.g., $f_{\nu_p}$, is in units of mJy, $\G= 300\times\G_{300}$ and the convention $Q= Q_n\times10^n$, e.g., $\nu= \nu_{19}\times10^{19}$ Hz and $B= B_5\times10^5$ Gauss, is used for other quantities; the same notations will be used in the rest of the paper. In the following, our discussion will be based on the synchrotron radiation only. But for the use of reference the expression of $\nu_a$ for SSC is also given here.

%%%%%%%%%%%%%%%%%%%%%%%%%%%%%%%%%%%%%%%%%%%%%%%%%%%%%%%%%%%%%%%%%%%%%%%%%%%%%%%%%

%%%%%%%%%%%%%%%%%%%%%%%%%%%%Begin Fig. 1%%%%%%%%%%%%%%%%%%%%%%%%%%%%%%%%%
\begin{figure*}
\centerline{\includegraphics[width=11.5cm,angle=0]{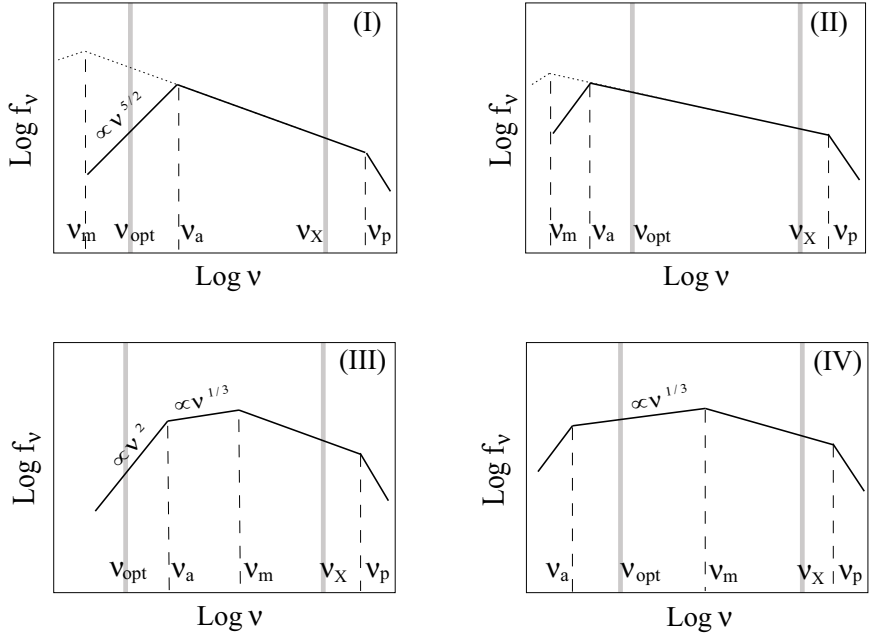}}
\caption{The four cases of the broad-band synchrotron spectrum of the
GRB prompt emission discussed in the text. The broken power law shape around
$\nu_p$ is phenomenologically derived from the observed $\g$-ray spectrum in
GRBs; see its possible theoretical origins in Sec. 2.
$\nu_a$ is the self absorption frequency. The dotted line in the two top
panels is the low energy part of the spectrum when the self absorption is
not considered. The two grey vertical bars marks the positions of
$\nu_{opt}$ and $\nu_X$, respectively. $\nu_X =$ 0.3 keV is the lower
end of the Swift XRT band pass.} \label{fig:spec_case}
\end{figure*}
%%%%%%%%%%%%%%%%%%%%%%%%%%%%End Fig. 1%%%%%%%%%%%%%%%%%%%%%%%%%%%%%%%%%%%

%%%%%%%%%%%%%%%%%%%%%%%%%%%%%%%%%%%%%%%%%%%%%%%%%%%%%%%%%%%%%%%%%%%%%%%%%%%%%%%%%
\section{The one-spectral-component assumption}

In this paper, we make an assumption that for the two samples (see Sec. 5 for a description of the sample selection criteria) studied,
the optical and $\g$/X-ray photons belong to a same synchrotron continuum spectrum generated by a same group of hot electrons. This assumption is based on the following three considerations. First, GRB prompt $\g$/X-ray emission is often interpreted as synchrotron radiation of a group of non-thermal electrons. If that is the case, the synchrotron spectrum {\it must} have a continuum extending to the low-frequency regime up to the optical band, presumably with a gentle slope of $f_{\nu} \propto \nu^{1/3}$ unless it has a self-absorption break. Secondly, the optical flux density is expected to always lie below or near the extrapolation from the $\g$-ray spectrum. This is generally consistent with observations in our sample (see Sec. 5 below and Fig. 2). Finally, a temporal correlation between the optical flux variation and the $\g$-ray LC is observed for the 4 GRBs in the our sample, suggesting the two components likely have the same dynamical origin. This is the major supporting evidence for our assumption.

We notice that there are other scenarios on prompt $\g$/optical emission that have been discussed in the literature. These include the synchrotron + SSC model (Kumar \& Panaitescu 2008; Racusin et al. 2008), the models invoke different emission radii for optical and $\g$-ray emissions (Li \& Waxman 2008; Fan, Zhang \& Wei 2009), and the model invokes two shock regions at a same emission radius (Yu, Wang \& Dai 2009). These models are relevant to GRB 080319B, which clearly requires a separate spectral component to interpret the prompt optical emission. For most other bursts studied in this paper, although these models are not ruled out, they are not demanded by the data. Our simple one-component model is adequate to interpret these bursts, and we will hereafter adopt this one-component assumption.

%%%%%%%%%%%%%%%%%%%%%%%%%%%%%%%%%%%%%%%%%%%%%%%%%%%%%%%%%%%%%%%%%%%%%%%%%%%%%%%%%%%%%

%%%%%%%%%%%%%%%%%%%%%%%%%%%%%%%%%%%%%%%%%%%%%%%%%%%%%%%%%%%%%%%%%%%%%%%%%%%%%%%%%%%%%
\section{Deriving $R$ constraints from $\nu_a$ and SED}
\label{sec:R-drv}

Let us consider the synchrotron emission as the prompt emission mechanism.
Depending on the locations of $\nu_m$ and $\nu_a$, the ratio of the optical flux density to the flux density at $\nu_p$ is
given for four different spectral cases by
\begin{displaymath}
\frac{f_{\nu_{opt}}}{f_{\nu_p}} =
\end{displaymath}
\begin{equation}
\left\{ \begin{array}{ll}
\left(\frac{\nu_a}{\nu_p}\right)^{\b_1} \left(\frac{\nu_{opt}}{\nu_a}\right)^{\frac{5}{2}},
& \!\!\!\!\! {\rm if \,\,} \nu_m < \nu_{opt} < \nu_a \,\, - {\rm \ case \, (I),}\\
\left(\frac{\nu_{opt}}{\nu_p}\right)^{\b_1},
& \!\!\!\!\! {\rm if \,\,} \nu_m < \nu_a < \nu_{opt} \,\, - {\rm \ case \, (II),}\\
\left(\frac{\nu_m}{\nu_p}\right)^{\b_1} \left(\frac{\nu_a}{\nu_m}\right)^{\frac{1}{3}} \left(\frac{\nu_{opt}}{\nu_a}\right)^2,
& \!\!\!\!\! {\rm if \,\,} \nu_{opt} < \nu_a < \nu_m \,\, - {\rm \ case \, (III),}\\
\left(\frac{\nu_m}{\nu_p}\right)^{\b_1} \left(\frac{\nu_{opt}}{\nu_m}\right)^{\frac{1}{3}},
& \!\!\!\!\! {\rm if \,\,} \nu_a < \nu_{opt} < \nu_m \,\, - {\rm \ case \, (IV).}\\
\end{array} \right.
\end{equation}

Fig. \ref{fig:spec_case} illustrates these four spectral cases. There is also a variation of case (III) (let us call it case III.5): $\nu_{opt} < \nu_m < \nu_a$, for which the flux density ratio is $f_{\nu_{opt}}/f_{\nu_p}= (\nu_a/\nu_p)^{\b_1} (\nu_m/\nu_a)^{5/2}(\nu_{opt}/\nu_m)^2$. We will come back to this case later and show that it is almost exactly the same as case (III). In some GRBs simultaneous observations of prompt X-ray and $\g$-ray emissions are made. The X-ray spectrum corrected for the photo-absorption in the soft end always nicely matches the power law extrapolated from $\g$-ray spectrum below $\nu_p$, without a need to invoke a break (e.g. Cenko et al. 2006; Romano et al. 2006; Page et al. 2007). This requires that both $\nu_a$ and $\nu_m$ be smaller than $\nu_X=$ 0.3 keV, the lower end of the Swift XRT band pass. We take this as a constraint in our analyses.

We aim to constrain $R$ based on the spectral information, such as $f_{\nu_p}$, $f_{\nu_{opt}}$ and $\b_1$, using $\nu_a$ as a proxy. We know from Sec. 2 that $\nu_a$ is expressed in terms of $f_{\nu_p}$, $\nu_m$ and $R$. So for each spectral case, we substitute the appropriate $\nu_a$ expression into the flux density ratio equation or into the constraint on $\nu_a$ implied by the definition of that spectral case (Eq. 8), and then get the $R$ constraints.

For case (I), substituting $\nu_a$ and letting $\nu_{opt}=5\times10^{14}$ Hz (for R band), we have
\begin{equation}
R_{14} = 7.5\times\left(\frac{C_2}{2}\right)^{\frac{1}{2}} f_{\nu_{opt}}^{\frac{1}{2}} \frac{D_{L,28}}{(1+z)} B_5^{\frac{1}{4}} \left(\frac{\G_{300}}{1+z}\right)^{\frac{3}{4}}.
\end{equation}
There is also a justification criterion due to the case definition $\nu_a > \nu_{opt}$, where $\nu_a$ is directly determined from the flux density ratio in Eq. (8) by
\begin{equation}
\nu_a= 10^{(19-\frac{12.2}{2.5-\b_1})} \left( \nu_{p, 19}^{-\b_1} \frac{f_{\nu_p}}{f_{\nu_{opt}}} \right)^{\frac{1}{2.5-\b_1}} \,\, {\rm Hz}.
\end{equation}

For case (II), the flux density ratio does not depend on $\nu_a$ and hence on $R$. One justification criterion for this case is that the spectral slope from the optical to the X- or $\g$-rays has to be consistent with $\b_1$, i.e., $\b_{opt-X/\g}= \b_1$. Another criterion due to the case definition is $\nu_a < \nu_{opt}$. Substituting with the expression of $\nu_a$, the latter gives
\begin{displaymath}
R_{14}  > 7.5\times\left(\frac{C_2}{2}\right)^{\frac{1}{2}} \times (2.2)^{\b_1} \times10^{-2.5\b_1}
\end{displaymath}
\begin{equation}
\hspace{2cm} \times \nu_{p,19}^{-\frac{\b_1}{2}} f_{\nu_p}^{\frac{1}{2}} \frac{D_{L,28}}{(1+z)} B_5^{\frac{1}{4}} \left(\frac{\G_{300}}{1+z}\right)^{\frac{3}{4}}.
\end{equation}

For case (III), when substituting the appropriate $\nu_a$ expression in Eq. (6) into the flux density ratio relation, it gives the expression of $R$ in
\begin{equation}
R_{14} = 2.0\times\left(\frac{C_1}{2}\right)^{\frac{1}{2}}
\nu_{m,17}^{-\frac{1}{4}} f_{\nu_{opt}}^{\frac{1}{2}} \frac{D_{L,28}}{(1+z)} B_5^{\frac{1}{4}} \left(\frac{\G_{300}}{1+z}\right)^{\frac{3}{4}}.
\end{equation}
We find for case (III.5) that the $R$ expression --- obtained by substituting the appropriate $\nu_a$ expression into Eq.(8) --- is exactly the same as Eq. (12) except that $C_1$ is replaced with $C_2$. We find the ratio $C_2/C_1$ lies in the range of (1, 1.6) for $\b_1= (-1.4, 0)$ (see in Appendix).

According to the definition of case (III), $\nu_{opt} < \nu_m < \nu_X$. Plugging this constraint of $\nu_m$ into Eq. (12) and its counterpart equation for case (III.5), it gives
\begin{displaymath}
2.1\times\left(\frac{C_1}{2}\right)^{\frac{1}{2}} f_{\nu_{opt}}^{\frac{1}{2}} \frac{D_{L,28}}{(1+z)} < R_{14} B_5^{-\frac{1}{4}} \left(\frac{1+z}{\G_{300}}\right)^{\frac{3}{4}} <
\end{displaymath}
\begin{equation}
\hspace{2.5cm} 7.5\times\left(\frac{C_2}{2}\right)^{\frac{1}{2}} f_{\nu_{opt}}^{\frac{1}{2}} \frac{D_{L,28}}{(1+z)}.
\end{equation}
Note that in writing this constraint we already combined the one for case (III) with the one for case (III.5). It is done by using $C_1$ in the lower limit and $C_2$ in the upper limit, such that the combined constraint is conservative. From now on we expand the case (III) definition to be $\nu_{opt} < \min(\nu_a, \nu_m)$ so that it includes case (III.5).

%%%%%%%%%%%%%%%%%%%%%%%%BEGIN Fig. 2%%%%%%%%%%%%%%%%%%%%%%%%%%%%%%%%%%%%%%%
\begin{figure*}
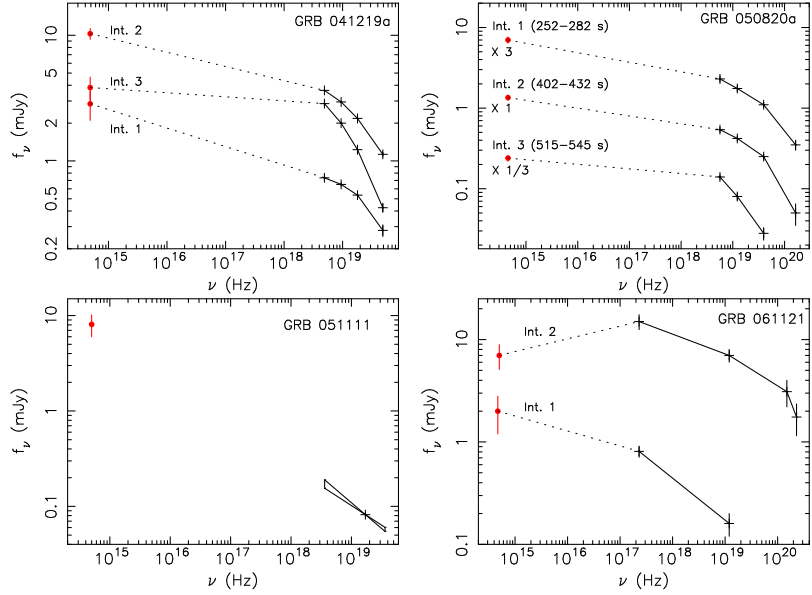

\centerline{
\includegraphics[width=3.9cm,angle=270]{041219a_spec_data.ps}
\hspace{0.1cm}
\includegraphics[width=3.9cm, angle=270]{050820a_spec_data.ps}
}
\centerline{
\includegraphics[width=3.9cm,angle=270]{051111_spec_data.ps}
\hspace{0.1cm}
\includegraphics[width=3.9cm, angle=270]{061121_spec_data.ps}
}
\caption{Observed optical to $\g$/X-ray broadband spectra for the 4 GRBs in our
prompt optical detection sample. The dotted line is a line connecting the
optical and the lowest energy $\g$/X-ray data points. For GRB 051111, only one
$\g$-ray data point is shown, superposed with the single-power-law fitted spectral
index and its confidence range.  Adapted from Vestrand et al. (2005),
Vestrand et al. (2006), Yost et al. (2007a) and Page et al. (2007), respectively.}
\label{fig:obs}
\end{figure*}
%%%%%%%%%%%%%%%%%%%%%%%%%END Fig. 2%%%%%%%%%%%%%%%%%%%%%%%%%%%%%%%%%%%%%%%%%

Two last pieces of constraining information for case (III) are from $\nu_a$, i.e., $\nu_a(f_{\nu_p}, \b_1, R)< \nu_X$ and $\nu_a(f_{\nu_p}, \b_1, \nu_m, R)$ $> \nu_{opt}$, where we use the expressions for $\nu_a$ given in Eq. (6); the first $\nu_a$ expression is for the situation of $\nu_m < \nu_a$ and the second for the situation of $\nu_a < \nu_m$. The first constraint gives
\begin{displaymath}
R_{14}B_5^{-\frac{1}{4}} \left(\frac{1+z}{\G_{300}}\right)^{\frac{3}{4}}> \left(\frac{C_2}{2}\right)^{\frac{1}{2}}\times(0.85)^{1+\b_1}
\end{displaymath}
\begin{equation}
\hspace{3cm} \times 10^{-1.75-\b_1} f_{\nu_p}^{\frac{1}{2}} \nu_{p,19}^{-\frac{\b_1}{2}} \left(\frac{D_{L,28}}{1+z}\right).
\end{equation}
The $\nu_a$ expression in the second constraint, $\nu_a > \nu_{opt}$, contains $\nu_m$ which can be expressed in terms of $f_{\nu_{opt}}$ and $R$ from the $R$-expression for this spectral case. After substituting for $\nu_m$, the second constraint gives
\begin{displaymath}
R_{14} B_5^{-\frac{1}{4}} \left(\frac{1+z}{\G_{300}}\right)^{\frac{3}{4}} > \left(\frac{C_1}{2}\right)^{\frac{1}{2}}\times2.0
\end{displaymath}
\begin{equation}
\hspace{0.5cm} \times \biggl[ 2.4\times 10^{\b_1} \left(\frac{f_{\nu_{opt}}}{f_{\nu_p}}\right)^{\frac{1}{2}}
\nu_{p,19}^{\frac{\b_1}{2}} \biggr]^{\frac{1}{2/3-2\b_1}}
f_{\nu_{opt}}^{\frac{1}{2}} \frac{D_{L,28}}{(1+z)}.
\end{equation}
Here the constraint $\nu_a > \nu_{opt}$ gives a lower limit of $R$, contrary to what is inferred from the conventional relation between $\nu_a$ and $R$. It is because in this subtle occasion $\nu_a$ depends not only on $R$ but also on $\nu_m$, and $\nu_m$ is expressed in terms of the optical flux density and $R$, hence the combined $R$-dependence of $\nu_a$ is positive.

The final constraint for case (III) should be the overlapping region among those three constraints obtained.

For case (IV), $\nu_m$ can be obtained from the flux density ratio relation by
\begin{equation}
\nu_{m,17} = \biggl[ 6\times10^{2\b_1} \left(\frac{f_{\nu_{opt}}}{f_{\nu_p}}\right)
\nu_{p,19}^{\b_1} \biggr]^{\frac{1}{\b_1-1/3}},
\end{equation}
whose value will be used to justify the case definition $\nu_{opt} < \nu_m < \nu_X$. Another constraint from the case definition is $\nu_a(f_{\nu_p}, \b_1, \nu_m, R) < \nu_{opt}$. Substituting with the $\nu_m$ expression, this gives
\begin{displaymath}
R_{14} B_5^{-\frac{1}{4}} \left(\frac{1+z}{\G_{300}}\right)^{\frac{3}{4}} > \left(\frac{C_1}{2}\right)^{\frac{1}{2}}\times 2.0
\end{displaymath}
\begin{equation}
\hspace{0.5cm} \times \biggl[ 2.4\times 10^{\b_1} \left(\frac{f_{\nu_{opt}}}{f_{\nu_p}}\right)^{\frac{1}{2}}
\nu_{p,19}^{\frac{\b_1}{2}} \biggr]^{\frac{1}{2/3-2\b_1}}
f_{\nu_{opt}}^{\frac{1}{2}} \frac{D_{L,28}}{(1+z)}.
\end{equation}
Notice that two contrary constraints, $\nu_a > \nu_{opt}$ in case (III) and $\nu_a < \nu_{opt}$ in case (IV), give exactly the same constraints on $R$. This is because the $\nu_a$ expression in both cases contains $R$ and $\nu_m$, but in case (III) $\nu_m$ is a strong function of $R$, $\propto R^{-4}$, while in case (IV) $\nu_m$ is a function of the flux density ratio only. Thus in case (III) the $R$-dependence is reversed between two sides of the inequality relation $\nu_a > \nu_{opt}$.

To summarize, the overall constraints on $R$ are: Eq. (9) for case (I), Eq. (11) for case (II), Eq. (13-15) for case (III) and Eq. (17) for case (IV). In addition, when they are available, the calculated $\nu_a$ or $\nu_m$ must satisfy the case definitions.

If the optical flux density has only an upper limit, the above $R$-constraints must be taken with a conservative point of view wherever $f_{\nu_{opt}}$ is involved.
Let $f_{\nu_{opt}}$ represent the measured upper limit. For case (I), Eq. (9) will give an upper limit for $R$. For case (II), Eq. (11) remains. For case (III), Eq. (13) is left with only the upper limit of $R$, Eq. (14) remains and Eq. (15) is useless. For case (IV), Eq. (16) gives a lower limit of $\nu_m$ which can be used to justify the case definition; Eq. (17) is useless. But we recall that Eq. (17) is obtained by substituting the $\nu_m$ expression into the definition constraint $\nu_a(f_{\nu_p}, \b_1, \nu_m, R) < \nu_{opt}$. Here, instead of using the $\nu_m$ expression, we plug in the upper boundary of $\nu_m$: $\nu_m < \nu_X$, then a new lower limit of $R$ is obtained for case (IV):
\begin{displaymath}
R_{14}B_5^{-\frac{1}{4}} \left(\frac{1+z}{\G_{300}}\right)^{\frac{3}{4}}> \left(\frac{C_1}{2}\right)^{\frac{1}{2}}\times 0.94\times (0.85)^{\b_1}
\end{displaymath}
\begin{equation}
\hspace{3cm} \times 10^{-\b_1} f_{\nu_p}^{\frac{1}{2}} \nu_{p,19}^{-\frac{\b_1}{2}} \left(\frac{D_{L,28}}{1+z}\right).
\end{equation}

%%%%%%%%%%%%%%%%%%%%%%%%%%%%%%%%%%%%%%%%%%%%%%%%%%%%%%%%%%%%%%%%%%%%%%

%%%%%%%%%%%%%%%%%%%%%%%%%%%%%%%%%%%%%%%%%%%%%%%%%%%%%%%%%%%%%%%%%%%%%%%%%%%%%
\section{GRB data sample}

Now we turn to the real GRB data to which our method developed above can be applied. First we construct a small sample of GRBs whose prompt optical emission is not only detected but is also variable and temporally correlated with the $\g$-rays. Excluding GRB 080319B that requires a new spectral component for optical emission, we identify four of GRBs in the sample, all belonging to the long-duration class. Three of them show complex fluctuations in their prompt $\g$-ray and optical LCs while the fourth is a single, smoothly peaked event, so we utilize multiple time intervals for each of the three. The emission properties of each time interval are listed in Tab. 1.

GRB 041219A is a very long ($T_{dur}\sim$ 500 s) burst and has multiple peaks in $\g$-ray LC. It has three optical detection intervals; the first two are correlated with the first $\g$-ray peak and the third with the second $\g$-ray peak (Vestrand et al. 2005).

GRB 050820A is a similar one except that it has denser optical temporal coverage. Its optical LC is decomposed into two components : a smooth component with fast rise and power-law decay, and a strongly variable component superposed on it (Vestrand et al. 2006). The smooth component is well accounted for by the early afterglow due to the GRB outflow interacting with the ambient medium. The variable component is found to correlate with the $\g$-ray peaks, suggesting it has the same origin as the $\g$-rays. In Tab. 1 the optical emission properties for this burst are for the residual optical component after subtracting the smooth component.

GRB 051111 has a single FRED (fast rise and exponential decay) peak in $\g$-rays lasting $\sim$ 90 s. The first optical observation starts at 30 s after the burst trigger when the $\g$-ray LC began to decay (Yost et al. 2007a). The prompt optical LC (before the $\g$-rays die off) decays more steeply than that of the later optical afterglow emission. The prompt optical emission has an excess above the back extrapolation of the later optical afterglow component. It has a decay slope statistically compatible with that of the $\g$-ray LC, and its flux density is also compatible with the spectral extrapolation of the $\g$-rays (Yost et al. 2007a). This is good evidence that the prompt optical excess has the same origin as the $\g$-rays. We use the flux density of the excess - not the total - optical emission in the first optical observation time interval.

GRB 061121 has two separate $\g$-ray peaks. The last peak was caught by XRT, UVOT and ROTSE, and it appears in LCs in all bands (Page et al. 2007). We use the emission properties of two time intervals of the last peak, one during the rising phase, and the other just at the peak.

We show the SEDs for all time intervals of the 4 GRBs in Fig. \ref{fig:obs} using the data adapted from their original publications. Two SEDs show almost no break between the optical and the $\g$-rays (corresponding to the theoretical spectrum case II) or a break very close to the optical. In all other cases, at least one break is needed between the optical and the $\g$-rays bands. The break(s) could be $\nu_a$, $\nu_m$, or both.

Besides this first data sample for optical detections, we also define a second data sample which is composed of those optically ``dark" GRBs during the prompt phase. This sample is adopted from Yost et al. (2007a), who reported the bursts whose prompt phase was observed by ROTSE but only upper limits on the optical flux were retrieved. Each burst has either a single or multiple time intervals of ROTSE exposure during the prompt phase. For bursts with multiple optical time intervals, we use the interval which has the smallest measurement error in the $\g$-ray flux density $f_{\nu_p}$ - usually the interval that has the brightest $\g$-ray flux. The only exception is GRB 061222A, for which three time intervals are used. This is because all the three intervals are located at the brightest part of the $\g$-ray LC. They all have small errors in $f_{\nu_p}$, and the instantaneous $\g$-ray spectral index $\b_1$ are available for all three intervals. The motivation of selecting this sample is the following: Even if there is no direct detection of prompt optical emission, one can speculate the existence of a prompt optical emission component that tracking the $\g$-ray LCs, which is the spectral extension of the prompt $\g$-ray spectrum into the optical band. The flux level of this component must be fainter than the upper limit set by the ROTSE observations. We want to check whether inferred $R$ constraints of this sample is consistent with the sample with optical detection, and whether the non-detection of optical emission of this sample is due to stronger synchrotron self absorption associated with a smaller $R$.

\setcounter{table}{1}
\begin{table*}
\includegraphics[width=22cm,angle=0]{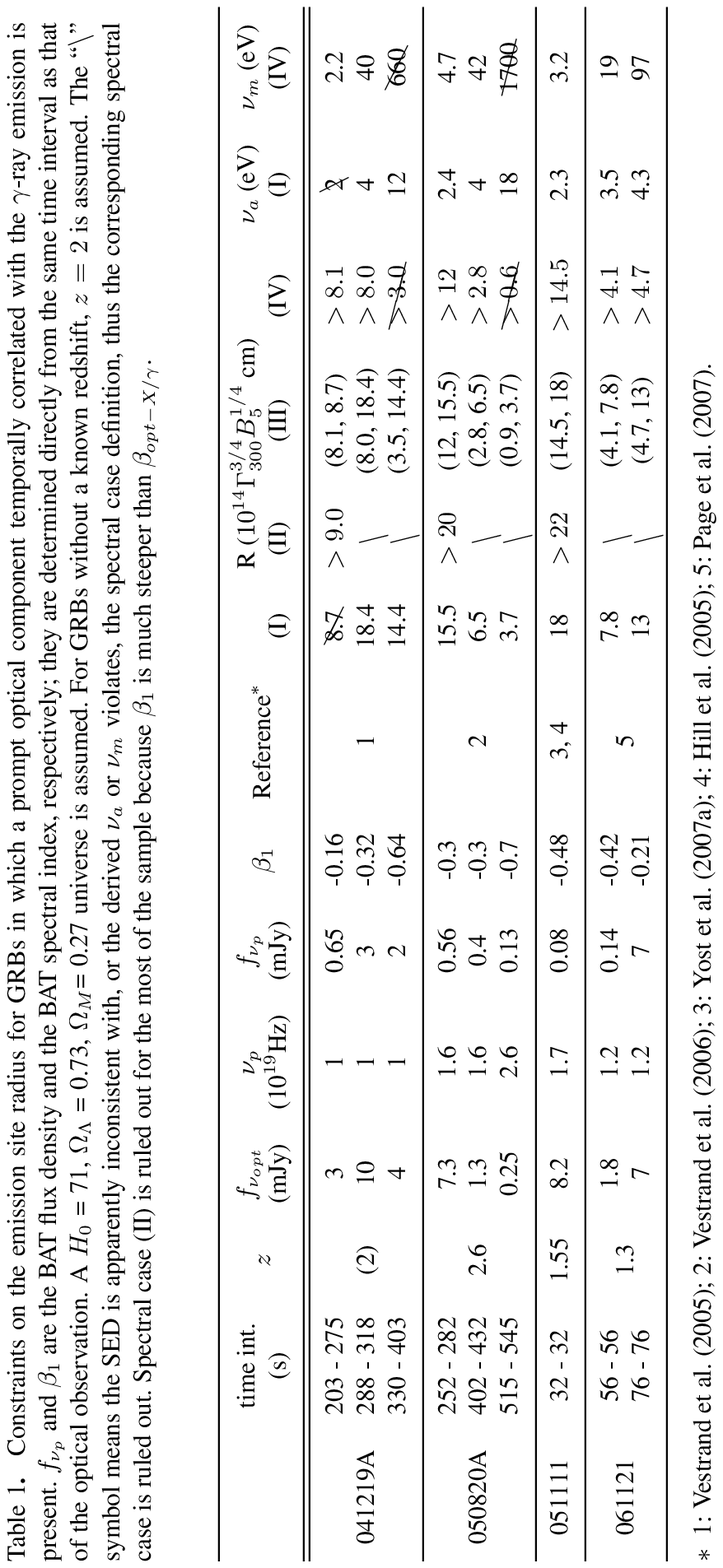}
\end{table*}
%%%%%%%%%%%%%%%%%%%%%%%%End Inserting Sideways Table%%%%%%%%%%%%%%%%%%%%%%%%

%%%%%%%%%%%%%%%%%%%%%%%%%%%%%%%%%%%%%%%%%%%%%%%%%%%%%%%%%%%%%%%%%%%%%%%%%%
\section{Results}

We apply the constraints on $R$ derived in Sec. \ref{sec:R-drv} to the first sample with prompt optical detections. The results for all four broad-band spectral cases are listed in Tab. 1. For most bursts in the sample, case (II) can be immediately ruled out because usually $\b_{opt-X} > \b_1$. Case (IV) is also ruled out for some bursts because the calculated $\nu_m$ is $\gg$ 0.3 keV. For GRB 041219A, case (I) can be ruled out for its first time interval because the case definition is not satisfied by the calculated $\nu_a$. Actually this interval is consistent with case (II), i.e., the optical intensity is consistent with the simple power-law extrapolation from the $\g$-ray spectrum.

\begin{figure*}
\centerline{\includegraphics[width=10cm,angle=0]{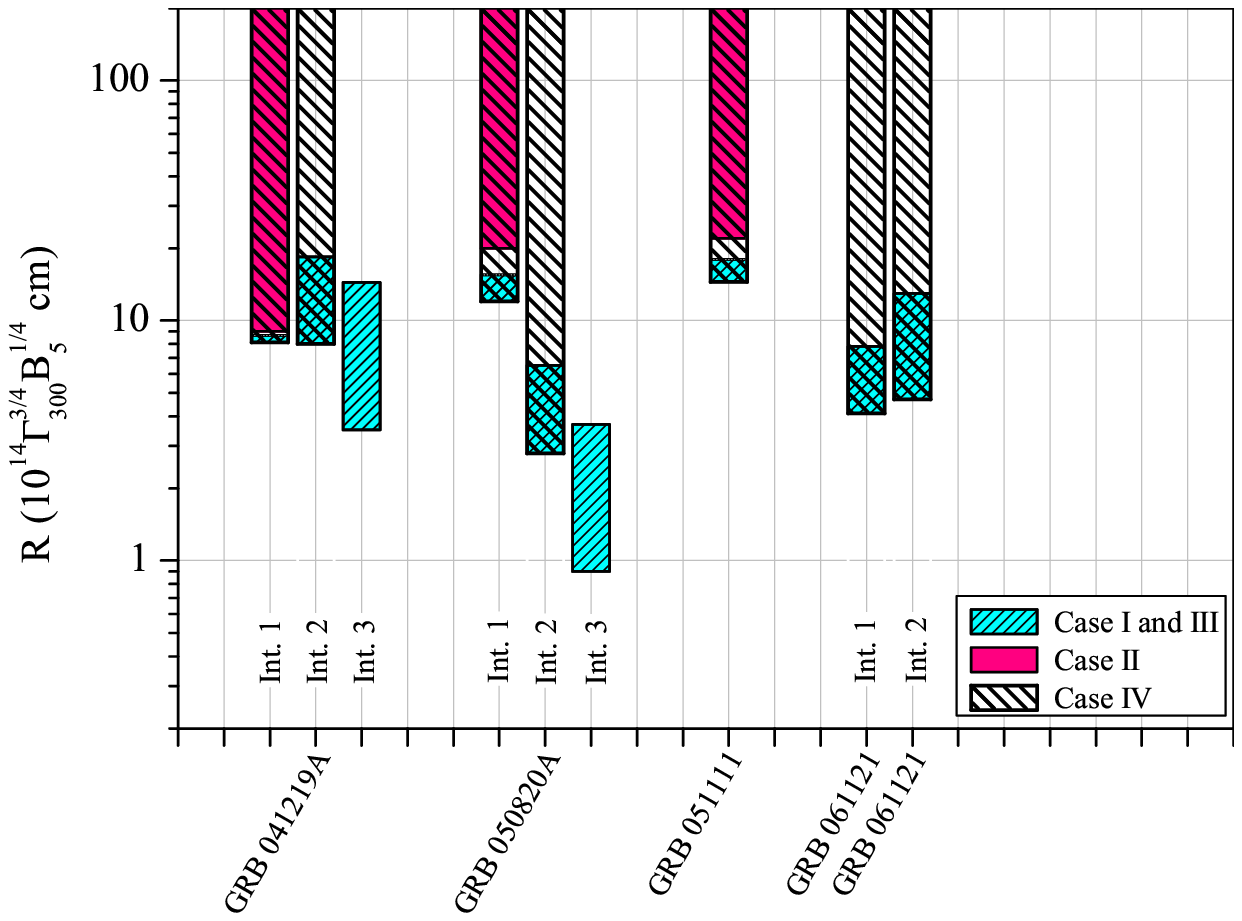}}
\caption{The constraints on the emission radii for 4 GRBs with prompt optical
detections. For 3 bursts of the sample, more than one time intervals per burst
are used. Based on the emission spectral information of individual bursts or time
intervals, spectral case II and IV, respectively, are already ruled out for miner
parts of the sample. See details in Tab. 1.} \label{fig:hist}
\end{figure*}

We plot the permitted $R$-ranges for each observation time interval in the sample for all possible spectral cases as floating bars in Fig. \ref{fig:hist}. The observed optical to $\g$-ray SED restricts $\nu_a$ from being much larger than $\nu_{opt}$. Accordingly, the results in Fig. \ref{fig:hist} give a constraint on the emission site for most time intervals of this sample: $R \geq$ a few $\times 10^{14}\,\,\G_{300}^{3/4} B_5^{1/4}$ cm. For two time intervals (041219A Int. 3 and 050820A Int. 3) in the sample, some spectral cases can be ruled out, thus the $R$-constraint can be pinned down to $R \approx (10^{14}-10^{15})\,\,\G_{300}^{3/4} B_5^{1/4}$ cm.

Similar results for the sample with only prompt optical upper limits are plotted in Fig. 4. In about half (6/13) of the sample a heavy self absorption, i.e. large $\nu_a$, is needed to account for the optical deficit, corresponding to the spectral case (I) and (III), which implies a constraint of $R < 10^{15}\,\,\G_{300}^{3/4} B_5^{1/4}$ cm. For the remaining half (7/13) of the sample, a spectral break at $\nu_m$ which is below $\nu_X$ but is much larger than $\nu_{opt}$ alone can give rise to the required deficit in optical, while $\nu_a$ can keep being smaller than $\nu_{opt}$, corresponding to case (IV). Thus for this half of the sample, we can provide no constraint on $\nu_a$ and hence on $R$.

Comparing Fig. 4 with Fig. \ref{fig:hist}, we find that there are always overlapping regions between the permitted $R$-ranges for the two samples. Therefore we can {\it not} draw any statistically significant distinction between these two samples as regards the constraints on their emission sites.

%%%%%%%%%%%%%%%%%%%%%%%%%%%%Begin Fig. 3%%%%%%%%%%%%%%%%%%%%%%%%%%%%%%%%%%%%%%%%
%%%%%%%%%%%%%%%%%%%%%%%%%%%%End Fig. 3%%%%%%%%%%%%%%%%%%%%%%%%%%%%%%%%%%%%%%%%%%

%%%%%%%%%%%%%%%%%%%%%%%%%%%%Begin Fig. 4%%%%%%%%%%%%%%%%%%%%%%%%%%%%%%%%%%%%%%%%
\begin{figure*}
\centerline{\includegraphics[width=10cm,angle=0]{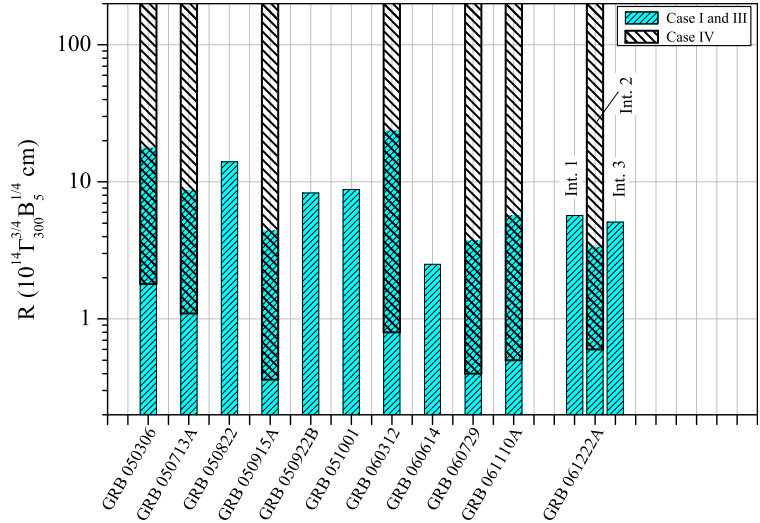}}
\caption{The constraints on the emission radii for GRBs without prompt optical detection. GRB 061222A has three time intervals that have information available for our calculation. Based on the emission spectral information, spectral case II is already ruled out for the whole sample, and spectral case (IV) is ruled out for about half of the sample. For 8 bursts without known redshift, $z=2$ is assumed.}
\end{figure*}
%%%%%%%%%%%%%%%%%%%%%%%%%%%%End Fig. 4%%%%%%%%%%%%%%%%%%%%%%%%%%%%%%%%%%%%%%%%%%

%%%%%%%%%%%%%%%%%%%%Begin Table 2  %%%%%%%%%%%%%%%%%%%%%%%%%%%%%%%%%%%%%%%%%%
\begin{table*}
\caption{The $\G$ constraints from the requirement that the photon annihilation optical depth $\tau_{\g\g} < 1$. A $H_0$ = 71, $\Omega_{\Lambda}$ = 0.73, $\Omega_M$= 0.27 universe is assumed. For the GRB without known redshift, $z=2$ is assumed.}
\label{tab:opa}
\begin{center}
\begin{tabular}{ll|llclllcc}
\hline
%%Entering the name for each column.
\multirow{2}{*}{GRB} & time int. & \multirow{2}{*}{$z$} & $\delta T$ & $N_1$ &  \multirow{2}{*}{$\al$} &
 $e_{max}$ & $e_{max, an}$ & \multirow{2}{*}{Reference$^{*}$} & \multirow{2}{*}{$\G$} \\
 & (s) & & (s) & (s$^{-1}$cm$^{-2}$MeV$^{-1}$) &  & (MeV) & ($\G_2^2$ MeV) & \\
\hline \hline
\multirow{3}{*}{041219A} & 203 - 275 & (2) & 15 & 0.15 & 1.80 & 0.2 & 1450 & \multirow{3}{*}{1} & $> 56$\\
 & 288 - 318 & (2) & 5 & 0.56 & 1.75 & 0.2 & 1450 & & $> 93$\\
 & 330 - 403 & (2) & 2 & 0.10 & 2.15 & 0.2 & 1450 & & $> 49$\\
\hline
\multirow{3}{*}{050820A} & 252 - 282 & 2.6 & 15 & 0.30 & 2.52 & 1.2 & 168 & \multirow{3}{*}{2} & $> 53$\\
 & 402 - 432 & 2.6 & 15 & 0.063 & 2.00 & 0.5 & 403 & & $> 53$\\
 & 515 - 545 & 2.6 & 10 & 0.015 & 1.96 & 0.25 & 806 & & $> 41$\\
\hline
\multirow{1}{*}{051111} & 32 - 32 & 1.55 & 8 & 0.035 & 1.48 & 0.15 & 2680 & 3 & $> 65$ \\
\hline
061121 & 76 - 76 & 1.3 & 1 & 2.7 & $< 2.9$ & 1.4 & 359 & 4 & $> 53$ \\
\hline
\end{tabular}
\\
\vspace{0.2cm}
\hspace{-3.3cm}
{\small $*$~~1: Vestrand et al. (2005); 2: Cenko et al. (2006); 3: Yost et al. (2007a); 4: Page et al. (2007).}
\end{center}
\end{table*}
%%%%%%%%%%%%%%%%%%%%%End Table 2 %%%%%%%%%%%%%%%%%%%%%%%%%%%%%%%%%%%%%%%%%%%%%

%%%%%%%%%%%%%%%%%%%%%%%%%%%%%%%Begin Tab. 3%%%%%%%%%%%%%%%%%%%%%%%%%%%%%%%%
\begin{table*}
\caption{The $\G$ constraints derived from the deceleration time $t_{dec}$ constraints for ISM and Wind medium, respectively. A $H_0$ = 71, $\Omega_{\Lambda}$ = 0.73, $\Omega_M$= 0.27 universe is assumed.
For the GRB without known redshift, $z=2$ is assumed. For GRB 041219A the inferred $t_{dec}$-constraint is very loose, partly because of a lack of X-ray afterglow observation, and also because the early infrared light curve ($t \leq 6\times 10^3$ s) is highly variable, possibly of internal shock origin, which makes it difficult to infer $t_{dec}$ to be these earlier times. In the cited reference for GRB 061121, only $E_{\g, iso}$ is given, without giving the $\g$-ray fluence. }
\label{tab:dec}
\begin{center}
\begin{tabular}{lclclcll}
\hline
%%Entering the name for each column.
 \multirow{2}{*}{GRB} & \multirow{2}{*}{$z$} & $t_{dec}$ & $\g$-ray fluence &
 $E_{\g, iso}$ & \multirow{2}{*} {Reference$^*$} & \multicolumn{2}{|c|} {$\G$}\\
 & & (s) & (10$^{-5}$ erg cm$^{-2}$) & (10$^{53}$ erg) & & ($\eta_{\g, 0.2}^{-1/8} n^{-1/8}$) & ($\eta_{\g, 0.2}^{-1/4} A_*^{-1/4}$)
 \\
\hline \hline
041219A & (2) & $< 3\times10^4$ & 15.5 & 15.2 & 1 & $> 73$ & $> 33$ \\
050820A & 2.6 & $< 500$ & 5.3 & 8.3 & 2, 3 & $> 337$ & $> 81$ \\
051111 & 1.55 & $< 100$ & 0.39 & 0.24 & 4 & $> 349$ & $> 46$ \\
061121 & 1.3 & $< 200$ & -- & 2.8 & 5 & $> 352$ & $> 70$ \\

\hline
\end{tabular}
\\
\vspace{0.2cm}
\hspace{-0.5cm}
{\small $*$~~1: Vestrand et al. (2005); 2: Vestrand et al. (2006); 3: Cenko et al. (2006); 4: Yost et al. (2007a); 5: Page et al. (2007).}
\end{center}
\end{table*}
%%%%%%%%%%%%%%%%%%%%%%%%%%%%%%%End Tab. 3%%%%%%%%%%%%%%%%%%%%%%%%%%%%%%%%%%

%%%%%%%%%%%%%%%%%%%%%%%%%%%%%%%%%%%%%%%%%%%%%%%%%%%%%%%%%%%%%%%%%%%%%%%%%%%
\subsection{The dependence of results on $\G$ and $B$}
\label{sec:Gam-B}

Strictly speaking, the constraint on $R$ is dependent on the source LF $\G$ and the magnetic field strength $B$. Independent determinations of $\G$ and $B$ for each GRB in our sample are not easy. So in this work we adopt the theoretically anticipated values in the standard internal-shock model: $\G \approx$ 300, $B \approx 10^5$ G. In the following we will justify these adopted values based on the available information of the GRBs in our sample.

\subsubsection{Constraints on $\G$}

Recently Molinari et al. (2007) inferred $\G \approx 400$ for two GRBs by directly observing the deceleration time of the GRB outflow. This value is in agreement with what we adopt. In addition, here we present some attempts to estimate $\G$ for each GRB in our first sample (with optical detections) using three independent arguments, which suggests that the choice of $\G \approx 300$ is reasonable.

(1) The variability time scale argument.
For the standard scenario in which the variability time scale is that of the central engine and that the ejecta form a conical jet with opening angle much larger than $1/\G$, the observed variability time scale should be at least the angular spreading time. This gives the constraint
\begin{equation}
\G > 41 [ R_{14} (1+z)/\delta T ]^{1/2},
\end{equation}
where $\delta T$ is the observed variability time scale, defined as the rising or
decaying time scale of the pulses in the LC. From the data we find that
$\delta T$ is $\approx$ 1 - 15 s for our sample. The $\G$-constraint from this argument is rather weak compared with the other two constraints described below.

(2) The photon annihilation opacity argument.
A GRB releases a huge amount of energy ($\sim 10^{53}$ erg isotropically) in terms of high energy photons from a small volume, which are subject to the photon-photon annihilation (e.g. Baring \& Harding 1997; Lithwick \& Sari 2001). Suppose $e_{max}$ is the maximum photon energy detected in the burst, the fact that a single power-law or a piece-wise power-law spectrum is detected for most GRBs up to $e_{max}$ implies that the optical depth of photon-photon pair production $\tau_{\g\g}$ for the photons of energy $e_{max}$ is smaller than unity.
$\tau_{\g\g}$ is related to the total number of photons and the size of the emission region, the latter of which can be expressed in terms of $\G$ and $\delta T$ within the internal shock model. Thus this requirement can impose a lower limit on $\G$ within the internal shock model. More generally the opacity argument can give a constraint in the $R-\Gamma$ space (Gupta \& Zhang 2008; Murase \& Ioka 2008; Zhang \& Pe'er 2009). Our following treatment applies to the internal shock model that is commonly discussed in the literature.

The minimum photon energy at which the photons are able to annihilate with photons of energy
$e_{max}$ is:
\begin{equation}
e_{max,an} = \frac{(\G m_e c^2)^2}{e_{max}(1+z)^2}.
\end{equation}
The power-law form spectrum just below $e_{max}$ is described as in
\begin{equation}
N(e)= N_1 \left(\frac{e}{\rm MeV}\right)^{-\al},
\end{equation}
where $e$ is the detected photon energy, $N(e)$ - in units of
$[$s$^{-1}$cm$^{-2}$MeV$^{-1}]$ - is the number of photons detected
per unit time per unit area per unit energy at $e$, $N_1$ is equal to
$N(e)$ at $e= 1$ MeV, and $\al$ is the photon index.

We followed the formulae of Lithwick \& Sari (2001) in their Limit A case to calculate
the lower limit of $\G$ due to $\tau_{\g\g} < 1$ for our optical detection sample.
The results, as well as the observational properties that are used, are summarized
in Tab. \ref{tab:opa}. Note that for all bursts in the sample, $e_{max, an} \gg
e_{max}$ for reasonable values of $\G$ (e.g., $\sim 10^2$). Thus our calculation has implicitly assumed that the power-law form spectrum detected below $e_{max}$ would actually extend well beyond $e_{max}$ and up to $e_{max, an}$, which is usually below the detector's limited bandpass. More rigorous calculations require the knowledge of the spectral shape below peak energy of the spectrum (Gupta \& Zhang 2008). However, the optical depth is much more sensitive to $\Gamma$ than the spectral indices, and the derived $\G$ constraint is not significantly modified by performing the more rigorous treatment. The above treatment is adequate to serve our purpose.

(3) The deceleration time constraint.
In the well established external shock model for GRB afterglows (e.g., M\'esz\'aros \& Rees 1997; Sari, Piran \& Narayan 1998; Chevalier \& Li 2000; see Piran 2005 for a review), the onset of afterglow marks the time, as known as the deceleration time $t_{dec}$, when one half of the total kinetic energy of the GRB outflow $E_k$ is deposited to the shocked circumburst medium. The afterglow light curve should rise before $t_{dec}$ and decay in a power law with time after $t_{dec}$ (e.g., Molinari et al. 2007). The deceleration time $t_{dec}$ is determined by $E_k$, outflow bulk LF $\G$ and the density of the circumburst medium, either a constant density medium (ISM) or a wind-like medium [$\rho(R) = A \times R^{-2}$]:
\begin{equation}
t_{dec} = \left\{ \begin{array}{ll}
\left(\frac{3 E_{k, iso}}{64\pi n m_p c^5 \G^8}\right)^{1/3} (1+z),
& {\rm \  for \ ISM},\\
\frac{E_{k, iso} (1+z)}{16\pi A c^3 \G^4},
& {\rm \  for \ Wind},\\
\end{array} \right.
\end{equation}
where $n$ is the proton number density of the ISM medium, and $A= 5\times 10^{11} A_*$ g cm$^{-1}$ is the Wind medium density normalization parameter. The isotropic equivalent kinetic energy $E_{k, iso}$ can be related to the isotropic energy release in $\g$-ray radiation $E_{\g, iso}$ by an energy conversion efficiency factor $\eta_{\g} = E_{\g,iso} / E_{k,iso}$.

The afterglow observations (X-rays and optical) for our sample show either a power law decay starting from the earliest observation interval or, a long-lasting shallow decay followed by a normal power law decay. It suggests that the afterglow onsets should be earlier than the start of the single power law decay or the start of the shallow decay. One can use the first observation data in the decaying afterglow phase to constraint the deceleration time to be earlier than the observational epoch (e.g., Zhang et al. 2006). From the data, we find the deceleration time $t_{dec} < 3\times10^4$ s for
GRB 041219a, and $t_{dec} <$ 100 - 500 s for the other 3 GRBs (050820a, 051111,
061121). Assuming $\eta_{\g} = 0.2$, $n= 1$ cm$^{-3}$ and $A_*= 1$, we find:
for ISM, $\G > 73$ for 041219a, $\G \gtrsim 350$ for the other 3 GRBs;
for Wind, $\G > 33$ for 041219a, $\G \gtrsim 50$ for the other 3. The results are
summarized in Tab. \ref{tab:dec}. We have checked of the compliance with closure relationships predicted by the external shock models for these bursts during the afterglow phase. It turns out that two (GRB 050820A and 061121) out of the 4 GRBs are consistent with and in favour of the ISM environment scenario, while the other two are consistent with both scenarios and can not discriminate between them.

To summarize, with the available data, one can only constrain but cannot determine $\G$ of the GRBs in our sample. On the other hand, all three constraints derived from the data are consistent with $\G = 300$ adopted in our calculations. In particular, the $\G$-constraint derived from the argument (3), which is the most stringent one among the three arguments, indicate that the assumed value of $\G = 300$ is reasonable.

\subsubsection{Constraints on $B$}

The $B$ value in the emission region is a function of $R$. There are two possible origins of the magnetic field in the emission region. The first component is the global magnetic field entrained by the ejecta from the central engine. Let's assume $B \sim 10^{14}$ G at the central engine, a typical value for a fast rotating magnetar or a fast rotating black hole accretion disk system - the two most plausible GRB central engine candidates. The $B$ value drops as $R^{-2}$ with $R$ up to the light cylinder, and then drops as $R^{-1}$ thereafter (Goldreich \& Julian 1969). Given that the central rotating source has a radius of $R_* \sim 10^6$ cm and a rotation period of $P\sim 1$ ms, at a radius $R \sim 10^{14}$ cm, the field that is carried within the outflow has a strength of $B \approx 2\times 10^5 B_{*,14}R_{*,6}^2 P_{ms}^{-1} R_{14}^{-1}$ G.

The second $B$ component is a random field generated {\em in-situ} in the emission region, likely in a relativistic shock via the Weibel instability (Medvedev \& Loeb 1999). This random field also follows the same $R$-dependence and is of the same order as the engine-related $B$ component if $\epsilon_B$ - the ratio of the post-shock magnetic energy density to the total energy density - is not too small (Zhang \& M\'{e}sz\'{a}ros 2002).

One can also briefly estimate the local field strength by relating the GRB $\g$-ray peak photon energy, typically $\sim 0.1$ MeV, with the synchrotron characteristic frequency, $\nu_p = \frac{eB \g^2}{2\pi m_e c}\frac{\G}{(1+z)}$, where $\g$ is the typical energy of electrons. In the internal-shock model, $\g= \epsilon_e f(p) (m_p/m_e) \theta_p$, where $\epsilon_e$ is the ratio of the electron energy density over the total thermal energy density in the post-shock fluid, $f(p)= (p-2)/(p-1)$ and $p$ is the electron energy spectral index. The parameter $\theta_p$ is the fractional energy gain of a proton passing the shock which depends only on the relative LF between the fluids downstream and upstream. For internal shocks, $\theta_p$ is {\it not} dependent on the shell bulk LF $\G$ and is of the order of unity. For $\epsilon_e = 0.3$ and $p=3$, we have $\g \approx 300$. Thus the required $B$ value can be estimated as $B \approx 5\times10^5 \G_{300}^{-1} \g_{300}^{-2} (\frac{1+z}{2})$ G.

The three crude estimates are marginally consistent with each other. We have taken $B=10^5 B_5$ G as the typical value throughout the text. Of course a large uncertainty exists due to our lack of understanding on the field properties, but it is reconciled by the very weak dependence of $R$ on $B$ (1/4 power).

%%%%%%%%%%%%%%%%%%%%%%%%%%%%%%%%%%%%%%%%%%%%%%%%%%%%%%%%%%%%%%%%%%%%%%%%%%%%%

\subsection{Comparison with results from an alternative modelling approach}

Kumar \& McMahon (2008) developed a general method of modelling GRB's $\g$-ray emission properties. Their method considers the  synchrotron and the SSC emission, respectively, as the radiation mechanism, takes into account the radiative cooling of electrons, and uses observed emission properties (such as the peak flux density and the pulse duration) to search for the allowed space of the model parameters such as $R$ and $\G$. Here we also apply their method for the synchrotron case to our optical detection sample using their code, and compare the results with ours. We add a new constraint into the module that controls the allowed model parameter space, which is that the optical flux density calculated from the model has to match the observed one within a $\pm 50$\% range.

To use this detailed modelling method, we have to specify which standard synchrotron spectral regime a GRB emission interval is in. The spectral indices ($\b_1$) of the optical detection sample in Tab. 1 have a variety of values around -1/2, based on which we classify the sample into 4 categories and apply the detailed modelling method accordingly. (1) For those time intervals that are most probably consistent with the $\b_1=-1/2$ regime (051111 and 061121 int. 1), this method gives $R \approx 10^{14} - 10^{16}$ cm. (2) For those possibly consistent with both $\b_1=-1/2$ and $\b_1=-(p-1)/2$ where $p>2$ (041219A int. 3 and 050820A int. 3), the method in the $\b_1=-1/2$ regime gives $R \approx 10^{14} - 10^{15}$ cm, while in the $\b_1=-(p-1)/2$ regime it gives no allowed $R$-space --- but if we relax the $f_{\nu_{opt}}$ constraint, it gives $R \approx 10^{17} - 10^{18}$ cm. (3) For those possibly consistent with both $\b_1=-1/2$ and $\b_1=-(p-1)/2$ where $1<p<2$ (041219A int. 2, 050820A int. 1 and 2), the method in the $\b_1=-1/2$ regime gives $R \approx 10^{14} - 10^{15}$ cm; however, the code provided by the authors is not applicable when $p < 2$. (4) The last category are those inconsistent with $\b_1=-1/2$ but probably consistent with $\b_1=-(p-1)/2$ if $1<p<2$ (041219A int. 1 and 061121 int. 2) for which the code is not applicable.

Overall, we find that for the major part of the optical detection sample where the detailed modelling method (Kumar \& McMahon 2008) is applicable the allowed spaces for $R$ from this method are about $10^{14} - 10^{16}$ cm. This is approximately consistent with the findings from our approach that $R$ is $\geq$ (a few $\times10^{14} - 10^{15})\,\,\G_{300}^{3/4} B_5^{1/4}$ cm for most of the intervals in the sample, and $10^{14}\,\,\G_{300}^{3/4} B_5^{1/4}~{\rm cm} < R < 10^{15}\,\,\G_{300}^{3/4} B_5^{1/4}$ cm for the remaining two intervals in the sample.

%%%%%%%%%%%%%%%%%%%%%%%%%%%%%%%%%%%%%%%%%%%%%%%%%%%%%%%%%%%%%%%%%%%%%%%%%%%

\section{Conclusion and Discussions}

Based on the assumption that the prompt optical and $\g$-ray emissions belong to the same synchrotron continuum of a group of hot electrons, we make constraints on the location of the prompt emission site for a sample of GRBs whose prompt optical emission is detected to temporally tracking the $\g$-ray LCs, by determining the location of $\nu_a$ in their SED. Our analysis shows that for most of the intervals in this sample the distance of the prompt emission site from the explosion centre $R$ is $\geq$ (a few $\times 10^{14}-10^{15})\G_{300}^{3/4} B_5^{1/4}$ cm, and for the remaining two intervals, the emission site is $(10^{14}-10^{15}) \G_{300}^{3/4} B_5^{1/4}$ cm away from the explosion centre.

The dependence of the distance constraint on the GRB outflow LF $\G$ is not negligible. On the other hand, various indirect observational constraints on $\G$ point to $\G \geq 300$ (e.g. Molinari et al. 2007; and even $\G \geq 600$ for GRB 080916C, Abdo et al. 2009). The derived observational constraints on $\G$ for bursts of this sample (Sec. \ref{sec:Gam-B}) are consistent with such an inference. In our paper, we take $\G = 300$ as a typical value. A higher $\G$ would only make the above distance constraint even larger. Our knowledge of the local magnetic field strength $B$ is less certain, although the $R$-dependence of $B$ is weak. Several crude estimates of $B$ based on the synchrotron radiation mechanism for GRB prompt emission suggest $B \sim 10^5$ G at $R\sim 10^{14}$ cm. This typical value has been adopted in our calculations.

The $R$-constraint we obtained is inconsistent with the photospheric emission model in which the prompt emission arises at the photosphere radius of $10^{11} - 10^{12}$ cm (Rees \& M\'{e}sz\'{a}ros 2005; Ryde et al. 2006; Thompson et al. 2007). This result alone can not discriminate between the fireball internal shock model and the magnetic outflow model. By comprehensive modeling the GRB prompt $\g$-rays and early X-rays, Kumar et al. (2007) concluded a prompt emission site of $R \sim 10^{15} - 10^{16}$ cm, which is supported by their further general modeling of the $\g$-ray emission properties (Kumar \& McMahon 2008). A large $R$ is derived for GRB 080916C through the pair opacity constraint (Abdo et al. 2009; Zhang \& Pe'er 2009)\footnote{Although both work obtained large $R$, the inference of $R$ in Abdo et al. (2009) is based specifically on the internal shock model while Zhang \& Pe'er (2009) gave a more model-independent constraint on $R$.} and for GRB 080319B through the synchrotron self-absorption constraint (Racusin et al. 2008) and the SSC scattering optical depth constraint (Kumar \& Panaitescu 2008). In summary, a large prompt emission distance from the central engine seems to be supported by three independent approaches, respectively, i.e. in $\g$/X-rays (e.g., Lazzati \& Begelman 2005; Lyutikov 2006; Kumar et al. 2007), in GeV $\g$-rays (Abdo et al. 2009; Zhang \& Pe'er 2009), and in the optical band (this paper).

We have also studied a sample of GRBs with prompt optical non-detections. Applying the same technique, we do {\it not} find any inconsistency between their $R$ constraints and those of the optical detection sample. This result is inherited from the findings by Yost et al. (2007b) that no distinction in distributions of $\b_1$ and $\b_{opt-\g}$ can be drawn between the optically dark GRBs and the GRBs with optical detections. However, this is only because the currently limited instrumental sensitivity prevents a distinction from being drawn. Deeper observations in the future in the optical band would provide further information regarding whether the optical deficit is due to a heavier synchrotron self-absorption in these bursts.

\subsection{Multi-color information for the low-energy spectrum}

If multi-color photometry near the optical band exists for the same time interval during the prompt phase, it would provide the local spectral index near the optical band (provided that the extinction correction is properly made). This would be helpful to identify the spectral case the data satisfy. For example, the spectral indices near optical differ by $\Delta \b = 5/3$ between the cases (III) and (IV), and by $\Delta \b = 1/2$ between the cases (I) and (III). Unfortunately, this kind of observational information is unavailable for all the time intervals of the optical detection sample we have considered in Tab. 1. The hope is that future multi-band prompt optical detections may be able to break the spectral case degeneracy and to tighten the $R$-constraint.

\subsection{Limitations of the method}

Our method is based on the assumption that the optical emission is emitted from the same group of electrons that produce the $\g$-rays in the same site via the same synchrotron radiation. This {\it one component} assumption has some supports (see Sec. 3) but certainly is not conclusive.

There are three other scenarios that have been discussed in the literature (mostly motivated by interpreting GRB 080319B). The first one invokes two emission zones for optical and $\g$-rays. For example, in the internal-shock-model based residual collision scenario proposed by Li \& Waxman (2008), the shells with high LF contrast in a GRB outflow collide first and merge at smaller radii, producing the $\g$/X-rays. Later those merged shells with low LF contrasts would collide mildly at later times and larger radii, giving rise to optical emission. Alternatively, if the outflow is neutron rich, the proton shells tend to collide at smaller radii to power $\g$/X-rays, while the free neutrons only decay at large radii and the decay products would be collided by later injected faster proton shells and power optical emission at larger radii (Fan et al. 2009). In both scenarios, it is expected that the observed optical pulse peak emission time is delayed by $R_{opt}/2\G^2 c$ with respect to the $\g$-ray pulse peak emission time, which may be in principle tested if the data quality is high. These models however do not naturally predict a smooth extension of $\g$-ray spectrum to the optical band without distinct spectral features. Although they cannot be ruled out by the data in our sample, they are more complicated than our one-zone model.

Secondly, our analysis is not applicable in the following scenario (Yu et al. 2009): a pair of shocks (reverse shock and forward shock) arise when two shells collide in the internal-shock model; different populations of electrons are accelerated in each of the two shocks and the two populations have different typical electron energies and different shock-generated magnetic strengths. The characteristic synchrotron frequencies are different - the forward shock produces the optical emission, while the reverse shock produces the $\g$-ray emission. These two emissions are two spectrally independent components but are temporally correlated because the heating of the two electron populations arises from the same dynamical process. Although this model may interpret the peculiar SED shape of GRB 080319B whose prompt optical flux density exceeds the extrapolation from the $\g$-ray spectrum by 4 orders of magnitude, it is unclear whether it can work properly for the bursts in our optical detection sample.

Finally, the synchrotron + SSC scenario (Kumar \& Panaitescu 2008; Racusin et al. 2008) has been proposed to interpret GRB 080319B. For our sample, there is no need to introduce a second distinct spectral component.
We consider only pure synchrotron radiation in deriving constraints on $R$, although we have given the $\nu_a$ estimation for SSC radiation whose value is similar to $\nu_a$ for synchrotron. In principle, it is possible that the observed emission from optical to $\g$-rays is dominated by SSC. If this is the case, our approach of constraining $R$ by calculating $\nu_a$ may give results somewhat different from the synchrotron case for the same sample of bursts. We did not carry out the analysis for the SSC case because the detailed shape of the SSC spectra is much more complicated than the synchrotron one. Note that our approach assumes that the optical and the $\g$-rays are from the same group of electrons due to the same radiation mechanism. In the SSC scenario this approach is applicable only if the SSC component dominates a large spectral band from the $\g$-ray down to the optical and the synchrotron component has to lie well below the optical band. This is usually not expected in the SSC models (e.g. Piran, Sari \& Zou 2009). In any case, if SSC is involved in interpreting any part of the spectrum in our sample, then our analysis is no longer applicable.

\section*{Acknowledgement}

We thank Pawan Kumar for valuable comments and suggestions and for sharing his code for our self-consistency check. We also acknowledge the anonymous referee whose comments have greatly helped to improve the quality and presentation of the paper. R-FS would like to thank Zhuo Li for useful comments and discussion. This work is supported by NASA grants NNG06GH62G and NNX07AJ66G (BZ) and by the NSF grant AST-0406878.

%%%%%%%%%%%%%%%%%%%%Begin the Reference%%%%%%%%%%%%%%%%%%%%%%%%%%%%%%%%%%%%%%%%

%%%%%%%%%%%%%%%%%%%%End the Reference%%%%%%%%%%%%%%%%%%%%%%%%%%%%%%%%%%%%%%%%%%

\renewcommand{\theequation}{A-\arabic{equation}}
% redefine the command that creates the equation no.
\setcounter{equation}{0}  % reset counter
\renewcommand{\thesubsection}{A. \arabic{subsection}}
% redefine the command that creates the subsection no.

\onecolumn

\section*{APPENDIX: Derivation of the self-absorption frequency}  % use *-form to suppress numbering

In this appendix, we provide a rigorous derivation of the blackbody equivalence equation (Eq. 3) that we use to calculate the self-absorption frequency (Eq. 3). The derivation is carried in the GRB ejecta comoving frame in which the relevant quantities are marked with the prime sign.

The self-absorption frequency $\nu_a'$ is defined as $\tau(\nu_a')= 1$, where $\tau(\nu_a')$ is the optical depth due to the self-absorption at $\nu'= \nu_a'$. The optical depth $\tau(\nu')= \int{\al_{\nu'}' ds'}$ deceases with the frequency, where $\al_{\nu'}'$ is the self-absorption coefficient [cm$^{-1}$Hz$^{-1}$] and the integral is over the line-of-sight width of the emitting source. The integral can be calculated directly only when we have the exact information on the number density of the emitting particles and its distribution over the length, which is not easy. Instead of directly calculating the integral $\int{\al_{\nu'}' ds'}$, we turn to derive the emission coefficient $j_{\nu'}'$ [erg s$^{-1}$cm$^{-3}$Hz$^{-1}$sr$^{-1}$] and then express the integral of $\al_{\nu'}'$ over width into the integral of $j_{\nu'}'$ over width, the latter is just the specific intensity at the source surface which is directly observable.

The synchrotron radiation spectrum, or specific radiation power, of an electron with LF $\g$ gyrating in a magnetic field $B$ with a pitch angle $\al$ is
\begin{equation}\label{power_syn}
P'(\nu', \g)= \frac{\sqrt{3}e^3 B \sin{\al}}{m_e c^2} F(\nu'/\nu_{ch}'),
\end{equation}
where $e$ is the electron charge and
\begin{equation}
\nu_{ch}'= \frac{3 e B \sin{\al} \g^2}{4\pi m_e c}
\end{equation}
is the characteristic photon frequency of the electron. The function
\begin{equation}
F(x)\equiv x \int_x^{\infty} K_{5/3}(\xi) d\xi
\end{equation}
has the asymptotic form
\begin{equation}
F(x) \sim \left\{ \begin{array}{lll}
\frac{4\pi}{\sqrt{3} \G(1/3)} (x/2)^{1/3} & \sim 2.15 x^{1/3}, & {\rm if} x \ll 1,\\ (\pi/2)^{1/2} e^{-x} x^{1/2} & \sim 1.25 e^{-x} x^{1/2}, & {\rm if} x \gg 1,
\end{array}\right.
\end{equation}
where $\G(1/3)$ is the gamma function of argument 1/3, and it reaches the maximum $F_{max}(x) \simeq 0.92$ at $x \simeq 0.29$. One integral property of the function $F(x)$ is
\begin{equation}\label{int_Fx}
\int_0^{\infty} x^{\mu} F(x) dx = \frac{2^{\mu+1}}{\mu+2} \G \left(\frac{\mu}{2}+\frac{7}{3}\right) \G \left(\frac{\mu}{2}+\frac{2}{3}\right),
\end{equation}
where $\G(y)$ is the gamma function of argument $y$. We will use this property later.

The self-absorption coefficient for any radiation mechanism is given (Rybicki \& Lightman 1979) by
\begin{equation}
\al_{\nu'}'= -\frac{1}{8\pi m_e\nu'^2}\int{d\g P'(\nu',\g)\g^2\frac{\partial}{\partial \g}\biggl[\frac{N(\g)}{\g^2}\biggr]},
\end{equation}
where $P'(\nu',\g)$ is the single electron's specific radiation power, and $N(\g)d\g$ is the number density of electrons with energy in the interval from $\g$ to ($\g+d\g$). In the case of GRB, $N(\g)$ has a two-power-law form and was described in Sec. 2 of the paper. We rewrite it here as
\begin{equation}
N(\g) = \left\{ \begin{array}{ll}
C_{\g} \g^{-p_1}, & {\rm if}\,\, \g_m < \g < \g_p \,,\\
C_{\g} \g_m^{(p_2-p_1)} \g^{-p_2}, & {\rm if}\,\, \g > \g_p \,,\\
\end{array} \right.
\end{equation}
where $C_{\g}$ is the normalization constant. Note that this distribution set-up is phenomenologically based on the two-power-law shape of the high energy radiation spectrum observed in GRBs; in the context of some specific particle acceleration scenario, e.g., shock acceleration, where two characteristic electron energies, i.e., the injection energy $\g_i$ and the cooling energy $\g_c$, are involved, there will be $\g_m=\min(\g_i, \g_c)$ and $\g_p=\max(\g_i, \g_p)$.
Thus $\partial [N(\g)/\g^2] / \partial \g =(-p-2)N(\g)/\g^3$, where $p$ could be either $p_1$ or $p_2$ depending on the location of $\g$. For the synchrotron radiation (Eq. \ref{power_syn}) the self-absorption coefficient would be
\begin{equation}\label{al_Fx}
\al_{\nu'}'= \frac{\sqrt{3}e^3 B \sin{\al} C_{\g}}{8\pi m_e^2 c^2 \nu'^2} \left[
(p_1+2) \int_{\g_m}^{\g_p} F(x) \g^{-(p_1+1)} d\g + (p_2+2) \int_{\g_p}^{\infty} F(x) \g^{-(p_2+1)} d\g \right],
\end{equation}
where $x \equiv x(\g) \equiv \nu'/\nu'_{ch}(\g)= (4\pi m_e c \nu')/(3e B\sin{\al} \g^2)$.

Then we consider two different locations of $\nu'$: $\nu' < \nu'_m$ and $\nu'_m < \nu' < \nu'_p$, respectively. If $\nu' < \nu'_m$, then $F(x)$ falls in the $\propto x^{1/3}$ asymptotic regime. One can transform the integral in Eq. (\ref{al_Fx}) for the variable $\g$ into the integral for the variable $x$. Notice the contribution from the second integral part in Eq. (\ref{al_Fx}) is unimportant, as long as $\g_m \ll \g_p$ and $1/3 < p_1 < p_2$. Thus it gives
\begin{equation}
\al_{\nu'}' = \frac{1}{2^{4/3}\G(1/3)} \frac{(p_1+2)}{(p_1+2/3)} \frac{e^3 B\sin{\al}C_{\g}} {m_e^2 c^2} \left(\frac{4\pi m_e c}{3e B\sin{\al}}\right)^{1/3}
\g_m^{-(p_1+2/3)} \nu'^{-5/3}.
\end{equation}

If $\nu'_m < \nu' < \nu'_p$, then $x(\g_m) \ll 1$ and $x(\g_p) \gg 1$. After the transformation of the variable $\g$ into the variable $x$, the first integral part in Eq. (\ref{al_Fx}) is in effect integrating over the $x$-range from $x(\g_m) \sim 0$ to $x(\g_p) \sim \infty$, thus we can use Eq. (\ref{int_Fx}) to calculate it. For the second integral part of Eq. (\ref{al_Fx}), $f(x) \propto x^{1/3}$, but its contribution is unimportant as long as $\g_m \ll \g_p$ and $1/3 < p_1 < p_2$. Therefore it gives
\begin{equation}
\al_{\nu'}'= \frac{\sqrt{3}e^3}{8\pi m_e^2 c^2} \left(\frac{3e}{2\pi m_e c}\right)^{p_1/2}C_{\g}(B\sin{\al})^{(p_1+2)/2}
\G\left(\frac{3p_1+22}{12}\right) \G\left(\frac{3p_1+2}{12}\right) \nu'^{-(p_1+4)/2}.
\end{equation}

To calculate $\nu_a'$ from $\tau(\nu_a')= \int{\al_{\nu_a'}ds'}=1$, one has to know the exact information about $N(\g_m)$ and its instantaneous distribution over the width of the emitting source along the line of sight, which are always subject to uncertainties. Nevertheless, in their attempts to calculate $\nu_a'$, some authors have calculated $N(\g_m)$ by assuming all electrons swept up by the shock are accelerated to relativistic energies either in the blastwave model for GRB afterglows (Granot, Piran \& Sari 1999; Panaitescu \& Kumar 2000; Pe'er \& Waxman 2004) or in the internal-shock model for prompt emissions (Li \& Waxman 2008). We warn that the reality in nature may be that {\it not all} but only a small fraction of the electrons encountered by the shock can be heated to relativistic energies and radiate, as was suggested by Bykov \& M\'{e}sz\'{a}ros (1996) and Daigne \& Mochkovitch (1998) (also see Kumar \& McMahon (2008) for an idea of repeated acceleration of a group of electrons), and this will introduce the biggest uncertainty to $N(\g_m)$ hence to this ``conventional'' approach of calculating $\nu_a'$. In general, this conventional approach over-estimates the number of emitting (and absorbing) electrons and, hence, over-estimates $\nu_a$. In the literature, it is usually suggested that $\nu_a$ is slightly below the X-ray band. According to our corrected calculation, $\nu_a$ is typically lower and can extend to close to the optical band in a wide parameter range.

Our new approach here is to express $\int{\al_{\nu'}ds'}$ in terms of $\int{j_{\nu'}'ds'}$, both of which contain the term $N
(\g_m)$ but the latter one is directly observable - it is just the specific intensity at the source surface. Therefore the new approach can avoid the uncertainties associated with partial acceleration and inhomogeneity over the source radial width.

Let us calculate the emission coefficient $j_{\nu'}'$. By definition,
\begin{equation}
4\pi j_{\nu'}'= \int_{\g_m}^{\g_p}{P'(\nu',\g)N(\g)d\g} + \int_{\g_p}^{\infty}{P'(\nu',\g)N(\g)d\g}.
\end{equation}
Following the same procedure of calculating $\al_{\nu'}'$, the integration gives
\begin{equation}
j_{\nu'}'= \left\{ \begin{array}{ll} \frac{1}{2^{1/3} \G(1/3)(p_1-1/3)}\frac{e^3 B\sin{\al} C_{\g}}{m_e c^2} \left(\frac{4\pi m_e c}{3e B\sin{\al}}\right)^{1/3} \g_m^{(1/3-p_1)} \nu'^{1/3}, & {\rm if \,\,} \nu' < \nu'_m,\\
\frac{2^{(p_1-1)/2} \sqrt{3}}{4\pi (p_1+1)} \frac{e^3 B\sin{\al} C_{\g}}{m_e c^2} \left(\frac{4\pi m_e c}{3eB\sin{\al}}\right)^{(1-p_1)/2} \G\left(\frac{3p_1+19}{12}\right) \G\left(\frac{3p_1-1}{12}\right) \nu'^{(1-p_1)/2}, & {\rm if \,\,} \nu'_m < \nu' < \nu'_p.\\ \end{array}\right.
\end{equation}

The ratio of $j_{\nu'}'$ over $\al_{\nu'}'$, also called the source function, is
\begin{equation} \label{source_func}
S_{\nu'}'= \frac{j_{\nu'}'}{\al_{\nu'}'}= \left\{ \begin{array}{ll} \frac{2(p_1+2/3)}{(p_1+2)(p_1-1/3)} m_e \g_m \nu'^2, & {\rm if \,\,} \nu' < \nu'_m,\\
\frac{\sqrt{2}}{p_1+1} \left(\frac{4\pi m_e c}{3e B\sin{\al}}\right)^{1/2} \frac{\G\left(\frac{3p_1+19}{12}\right) \G\left(\frac{3p_1-1}{12}\right)}{\G\left(\frac{3p_1+22}{12}\right) \G\left(\frac{3p_1+2}{12}\right)} m_e \nu'^{5/2}, & {\rm if \,\,} \nu'_m < \nu' < \nu'_p,\\ \end{array}\right.
\end{equation}
which does not have dependence on $N(\g_m)$. It shows that, for synchrotron radiation, the power-law index of the optical thick (to the self absorption) part of the emergent spectrum below $\nu_m'$ is 2, while the power-law index of the optical thick spectrum above $\nu_m'$ is 5/2.

From the definition of the self-absorption frequency $\int{\al_{\nu_a'}ds'}=1$, we have $\int{(j_{\nu_a'}'/S_{\nu_a'}') ds'}= 1$. Since $S_{\nu_a'}'$ does not depend on $N(\g_m)$ and its distribution over the source width, it can be taken out of the integral. Thus we have $S_{\nu_a'}'= \int{j_{\nu_a'}'ds'}= F_{\nu_a'}'$, where $F_{\nu_a'}'$ is the specific flux at the source surface in the asymptotic optically thin regime at $\nu_a'$ . Rewriting the expression for $S_{\nu'}'$ (Eq. \ref{source_func}) at $\nu'=\nu_a'$ and using the photon frequency vs. electron energy relation $\nu_{ch}'(\g)$ for synchrotron radiation, we get
\begin{equation}
\max(\g_m, \g_a)\times2m_e \nu_a'^2=  F_{\nu_a'}' C(p_1),
\end{equation}
where $\g_a$ is the energy of the electron whose characteristic photon frequency is $\nu_a'$, and the correction factor
\begin{equation}
C(p_1)= \left\{ \begin{array}{ll}
C_1(p_1)= \frac{(p_1+2)(p_1-1/3)}{p_1+2/3}, & {\rm if \,\,} \nu_a' < \nu'_m, \\
C_2(p_1)= \sqrt{2}(p_1+1) \frac{\G\left(\frac{3p_1+22}{12}\right) \G\left(\frac{3p_1+2}{12}\right)}{\G\left(\frac{3p_1+19}{12}\right) \G\left(\frac{3p_1-1}{12}\right)}, & {\rm if \,\,} \nu'_m < \nu_a' < \nu'_p. \\ \end{array}\right.
\end{equation}

If we assume a temperature $T'= \max(\g_m, \g_a)m_e c^2/k$ then the last equation is
\begin{equation}\label{kT_F}
2kT' \frac{\nu_a'^2}{c^2}= F_{\nu_a'}' C(\b_1),
\end{equation}
where for practical uses the correction factor $C(p_1)$ is changed to $C(\b_1)$ using the relation $\b_1= -(p_1-1)/2$, and so
\begin{equation}
C(\b_1)= \left\{ \begin{array}{ll}
C_1(\b_1)= \frac{(3-2\b_1)(2/3-2\b_1)}{5/3-2\b_1}, & {\rm if \,\,} \nu_a' < \nu'_m, \\
C_2(\b_1)= 2\sqrt{2}(1-\b_1) \frac{\G\left(\frac{25-6\b_1}{12}\right) \G\left(\frac{5-6\b_1}{12}\right)}{\G\left(\frac{11-3\b_1}{6}\right) \G\left(\frac{1-3\b_1}{6}\right)}, & {\rm if \,\,} \nu'_m < \nu_a' < \nu'_p. \\ \end{array}\right..
\end{equation}
In the samples presented in the main body of the paper, $\b_1$ is among -1.4 to 0, so the ranges for the correction factor are $C_1(\b_1)= (1.2, 4.5)$ and $C_2(\b_2)= (1.2, 7.0)$. Therefore Eq. (\ref{kT_F}) shows that, within a factor of a few, at $\nu_a'$ the un-absorbed source surface flux density is equal to the flux density of the Rayleigh-Jeans part of the blackbody spectrum with a temperature corresponding to the lowest energy of those electrons that are barely affected by the self absorption. This equation is used to calculate $\nu_a'$ in the main body of the paper where the correction factor $C(\b_1)$ is taken into account.
%%%%%%%%%%%%%%%%%%%%%%%%%%%%%%%%%%%%%%%%%%%%%%%%%%%%%%%%%%%%%%%%%%%%%%%%%%%%%%%%%%%%%%%%%%%%%%

%%%%%%%%%%%%%%%%%%%%%%%%%%%%%%%%%%%%%%%%%%%%%%%%%%%%%%%%%%%%%%%%%%%%%%%%%%%%%%%%%

\end{document}